\def\selectedoptions{}
\ifx\empty\selectedoptions
  \def\selectedoptions{final}
\fi

\documentclass[
   \selectedoptions
  ]
  {aipproc}

\def\selectedlayoutstyle{6x9} 
\layoutstyle\selectedlayoutstyle

\SetInternalRegister\hbadness{8000} % pseudo latin isn't breaking very well :-)

\newcommand\doingARLO[2][]{%
  \ifx\mmref\undefined #1\else #2\fi
}

\begin{document}
  \vspace*{-2.3cm}
  \begin{flushright}
  \noindent hep-ph/0112062 \\
  ANL-HEP-CP-01-118 \\
  \end{flushright}
%  \vspace*{6.84cm}
\title 
      []
      {Supersymmetry Explanation for the Puzzling Bottom Quark Production Cross Section}


\author{Edmond L. Berger}{
% If you represent a collaboration please use this format
%\author{Author representing the So-and-so collaboration}{
  address={High Energy Physics Division, Argonne National Laboratory, Argonne, IL 60439},
  email={berger@anl.gov},
  thanks={}
}

%\fi

% \copyrightholder{Acoustical Society of America}
\copyrightyear  {2001}

\begin{abstract}
It has been known for a long time that the cross section for bottom-quark $b$ 
production at hadron collider energies exceeds theoretical expectations.  An 
additional contribution from pair-production of light gluinos $\tilde{g}$, of mass 
12 to 16 GeV, with two-body decays into bottom quarks and light bottom squarks 
$\tilde{b}$, helps to obtain a $b$ production rate in better agreement with data.  
The masses of the $\tilde{g}$ and $\tilde{b}$ are restricted further 
by the ratio of like-sign to opposite-sign leptons at hadron colliders.  
Constraints on this scenario from other data are examined, and predictions
are made for various processes such as Upsilon decay into $\tilde{b}'s$.  
\end{abstract}

\date{\today}

\maketitle

\section{Introduction}
The cross section for bottom-quark production at hadron collider energies exceeds 
the central value of predictions of next-to-leading order (NLO) perturbative quantum 
chromodynamics (QCD) by about a factor of
two~\cite{expxsec,ua1data}.  This longstanding discrepancy has resisted fully 
satisfactory resolution within the standard model~\cite{qcdrev}.  The NLO 
contributions are large, and it is not excluded that a combination of further 
higher-order effects in production and/or fragmentation 
may resolve the discrepancy.  However, the disagreement is surprising because 
the relatively large mass of the bottom quark sets a hard scattering scale at
which fixed-order perturbative QCD computations of other processes are generally
successful.  The photoproduction cross section at DESY's HERA~\cite{herab} and 
and the cross section in photon-photon reactions at CERN's LEP~\cite{lepb} also 
exceed NLO expectations. 
The data invite the possibility of a contribution from ``new physics".  

\section{Supersymmetry Interpretation}
The properties of new particles that can contribute significantly to the bottom 
quark cross section are fairly well circumscribed.  To be produced with enough 
cross section the particles must interact strongly and have relatively low mass.  
They must either decay into $b$ quarks or be close imitators of $b$'s in a 
variety of channels of observation.  They must evade constraints 
based on precise data from measurements of $Z^o$ decays at CERN's LEP and SLAC's 
SLC, and from many lower-energy $e^+ e^-$ collider experiments.  The minimal 
supersymmetric standard model (MSSM) is a favorite candidate in many quarters for 
physics beyond the standard model.  It offers a well-motivated theoretical 
framework and is reasonably well-explored phenomenologically.  An explanation 
within the context of the MSSM~\cite{ourletter} can satisfy all of the stated 
criteria.  

In Ref.~\cite{ourletter}, 
the existence is assumed of a relatively light color-octet gluino 
$\tilde g$ (mass $\simeq 12$ to 16 GeV) that decays with 100\% branching 
fraction into a bottom quark $b$ and a light color-triplet bottom squark 
$\tilde b$ (mass $\simeq 2$ to 5.5 GeV).  The $\tilde g$ and the $\tilde b$ are 
the spin-1/2 and spin-0 supersymmetric partners of the gluon ($g$) and bottom 
quark.  In this scenario the $\tilde b$ is the lightest SUSY particle, and the 
masses of all other SUSY particles are arbitrarily heavy, i.e., of order the 
electroweak scale or greater.  The $\tilde b$ is either long-lived or decays via 
R-parity violation into a pair of hadronic jets.  Improved agreement is obtained 
with hadron collider rates of bottom-quark production, and several predictions are 
made that can be tested readily with forthcoming data. 
 
\subsection{Differential Cross Section}
The light gluinos are produced in pairs via standard QCD subprocesses, dominantly 
$g + g \rightarrow \tilde g + \tilde g$ at Tevatron and Large Hadron Collider (LHC) 
energies.  The $\tilde g$ has
a strong color coupling to $b$'s and $\tilde b$'s and, as long as its mass
satisfies $m_{\tilde g} > m_b + m_{\tilde b}$, the $\tilde g$ decays
promptly to $b + \tilde b$.  The magnitude of the $b$ cross section, the
shape of the $b$'s transverse momentum $p_{Tb}$ distribution, and the CDF
measurement~\cite{cdfmix} of $B^0 - \bar B^0$ mixing are three features of
the data that help to establish the preferred masses of the $\tilde g$ and
$\tilde b$.  In Ref.~\cite{ourletter}, contributions are included from both 
$q + \bar q \rightarrow \tilde g + \tilde g$ and 
$g + g \rightarrow \tilde g + \tilde g$.  The subprocess 
$g + b \rightarrow \tilde{g} + \tilde{b}$ contributes insignificantly.  Shown in 
Fig.~1 is the integrated $p_{Tb}$ distribution of the $b$ quarks that results
from $\tilde g \rightarrow b + \tilde b$, for $m_{\tilde g} = $14 GeV and
$m_{\tilde b} =$ 3.5 GeV.  The results are compared with the cross section
obtained from next-to-leading order (NLO) perturbative QCD
and CTEQ4M parton distribution functions (PDF's)~\cite{cteq}, with $m_b =$
4.75 GeV, and a renormalization and factorization scale $\mu=\sqrt{m_b^2 +
p_{Tb}^2}$.  SUSY-QCD corrections to $b \bar{b}$ production are not
included as they are not available and are generally expected to be
somewhat smaller than the standard QCD corrections.  A fully differential
NLO calculation of $\tilde g$-pair production and decay does not exist
either.  Therefore, the $\tilde g$-pair cross section is computed from the
leading order (LO) matrix element with NLO PDF's \cite{cteq},
$\mu=\sqrt{m^2_{\tilde{g}} + p^2_{T_{\tilde{g}}}}$, and a two-loop 
expression for  the strong coupling $\alpha_s$. To account for NLO effects, 
this $\tilde g$-pair cross section is multiplied by 1.9, the ratio 
of inclusive NLO to LO cross sections~\cite{prospino}.

A relatively light gluino is necessary in order to obtain a bottom-quark
cross section comparable in magnitude to the pure QCD component.  Values of
$m_{\tilde g} \simeq$ 12 to 16 GeV are chosen because the resulting $\tilde
g$ decays produce $p_{Tb}$ spectra that are enhanced primarily in the
neighborhood of $p_{Tb}^{\rm min} \simeq m_{\tilde g}$ where the data show
the most prominent enhancement above the QCD expectation.  Larger values of
$m_{\tilde g}$ yield too little cross section to be of interest, and
smaller values produce more cross section than seems tolerated by the ratio
of like-sign to opposite-sign leptons from $b$ decay, as discussed below.
The choice of $m_{\tilde b}$ has an impact on the kinematics of the $b$.
After selections on $p_{Tb}^{\rm min}$, large values of $m_{\tilde b}$
reduce the cross section and, in addition, lead to shapes of the $p_{Tb}$
distribution that agree less well with the data.  

\begin{figure}
\caption{Bottom-quark cross section in $p\bar p$ collisions at $\sqrt{S}
=1.8$ TeV for $p_{Tb}>p_{Tb}^{\rm min}$ with a gluino of mass
$m_{\tilde{g}} = 14$ GeV and a bottom squark of mass $m_{\tilde{b}} = 3.5$
GeV.  The dashed curve is the central value of the NLO QCD prediction. The
dotted curve shows the $p_T$ spectrum of the $b$ from the SUSY processes.  The 
solid curve is the sum of the QCD and SUSY components.  The shaded band 
represents an uncertainty of roughly $\pm$30\% associated with variations of 
the renormalization and factorization scales, the $b$ mass, and the parton 
densities.  The rapidity cut on the $b$'s is $|y_b| \le 1$. Data are from Ref.~1.}
\includegraphics[height=.5\textheight]{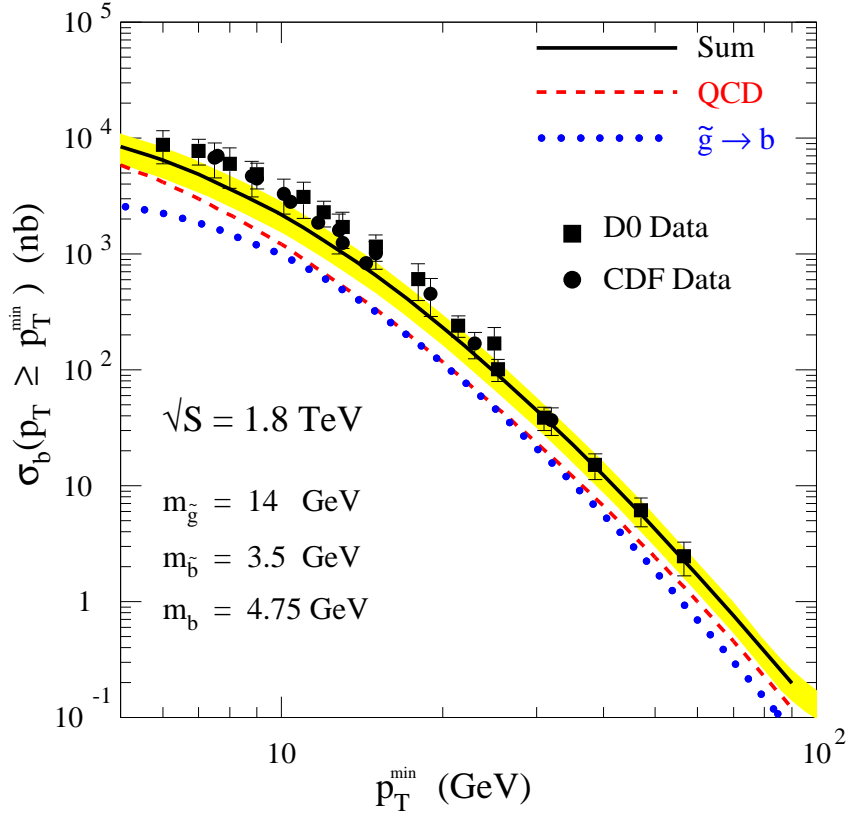}
\end{figure}

After the contributions of the NLO QCD and SUSY components are added, 
the magnitude of the bottom-quark cross section and the shape of the integrated 
$p^{\rm min}_{Tb}$ distribution are described well. Very good agreement is 
obtained also with data from the UA1 experiment (not shown)~\cite{ua1data}.
The SUSY process produces bottom quarks in a four-body final state and thus
their momentum correlations are different from those of QCD.  Angular
correlations between muons that arise from decays of $b$'s have been
measured \cite{cdfmix,muonexp}.  The angular correlations between
$b$'s in the SUSY case are nearly indistinguishable from those
of QCD once experimental cuts are applied.    

The energy dependence of the bottom cross section is a potentially
important constraint on models in which new physics is invoked to
interpret the observed excess bottom quark yield.  Since the 
assumed $\tilde{g}$ mass is larger than the mass
of the $b$, the $\tilde{g}$ pair process will turn on more slowly with
energy than pure QCD production of $b \bar{b}$ pairs.  The new physics
contribution will depress the ratio of cross sections at 630 GeV and
1.8 TeV from the pure QCD expectation.  An explicit calculation with 
CTEQ4M parton densities and the $b$ rapidity selection $|y| < 1$, yields 
a pure QCD prediction at NLO of 0.17 +/- 0.02 for 
$p_{Tb}^{\rm min} =$ 10.5 GeV, and 0.16 +/- 0.02 
after inclusion of the gluino pair contribution.  Either of these numbers 
is consistent with forthcoming data from CDF on this ratio~\cite{privcom}.  

\subsection{Same-sign to Opposite-sign Leptons}
Since the $\tilde g$ is a Majorana particle, its decay yields both quarks
and antiquarks.  Gluino pair production and subsequent decay to $b$'s will
generate $b b$ and $\bar b \bar b$ pairs, as well as the $b \bar b$
final states that appear in QCD production.  When a gluino is highly 
relativistic, its helicity is nearly the same as its chirality.  Therefore, 
selection of $\tilde g$'s whose transverse momentum is greater than their mass 
will reduce the number of like-sign $b$'s.  In the intermediate $p_T$ region,
however, the like-sign suppression is reduced.  The cuts chosen in current 
hadron collider experiments for measurement of the ratio of like-sign to 
opposite-sign muons result in primarily unpolarized $\tilde g$'s, and, 
independent of the
$\tilde b$ mixing angle, an equal number of like-sign and opposite-sign
$b$'s is expected at production.  The SUSY mechanism leads therefore to an 
increase of like-sign leptons in the final state after semi-leptonic decays of 
the $b$ and $\bar b$ quarks.  This increase could be confused with an enhanced 
rate of $B^0-\bar B^0$ mixing.  

Time-integrated mixing analyses of lepton pairs 
observed at hadron colliders are interpreted in terms of the 
quantity $\bar{\chi} = f_d \chi_d + f_s \chi_s$, where 
$f_d$ and $f_s$ are the fractions of
$B^0_d$ and $B^0_s$ hadrons in the sample of semi-leptonic $B$
decays, and $\chi_f$ is the time-integrated mixing probability for $B^0_f$.
Conventional $b\bar b$ pair production determines the quantity 
$LS_c = 2\bar{\chi} (1-\bar{\chi})$, the fraction of $b\bar b$ pairs that 
decay into like-sign leptons.  The SUSY mechanism leads to a new 
expression  
\begin{equation}
LS =\frac{1}{2} \frac{\sigma_{\tilde{g}\tilde{g}}}{\sigma_{\tilde{g}\tilde{g}}+
\sigma_{\rm{qcd}}} + 
LS_c \frac{\sigma_{\rm {qcd}}}{\sigma_{\tilde{g}\tilde{g}}+\sigma_{\rm {qcd}}} = 
2 \bar{\chi}_{\rm {eff}} (1 - \bar{\chi}_{\rm {eff}}).  
\end{equation}
The factor $1/2$ arises because $N(bb + \bar{b}\bar{b}) \simeq N(b \bar{b})$ in 
the SUSY mechanism for the selections on $p_{Tb}$ made in the CDF run I analysis.  
Defining $G = \sigma_{\tilde{g}\tilde{g}} / \sigma_{\rm{qcd}}$, 
the ratio of SUSY and QCD bottom-quark cross sections after cuts, and solving for 
the effective mixing parameter, one obtains
\begin{equation}
\bar{\chi}_{\rm {eff}}=\frac{\bar{\chi}}{\sqrt{1+G}} +{1\over 2}\left[1-
	\frac{1}{\sqrt{1+G}}\right] .
\end{equation}
The CDF measurement is interpreted in Ref.~\cite{ourletter} as a determination of 
$\bar{\chi}_{\rm {eff}} = 0.131 \pm 0.02 \pm 0.016$~\cite{cdfmix}.  This number  
is marginally larger than the world average value $\bar{\chi} = 
0.118 \pm 0.005$ \cite{pdg}, assumed to be the contribution from the pure QCD 
component only.  

The ratio $G$ is determined in the region of phase space where the measurement is 
made~\cite{cdfmix}, with both final $b$'s having $p_{Tb} \ge 6.5$ GeV and rapidity 
$| y_b | \leq 1$.  With $m_{\tilde b} =$ 3.5 GeV, $G =$ 0.37 and 0.28 for gluino 
masses $m_{\tilde g} =$ 14 and 16 GeV, respectively.  The predictions are 
$\bar{\chi}_{\rm {eff}} = 0.17 \pm 0.02 $ for $m_{\tilde g} =$ 14 GeV, and 
$\bar{\chi}_{\rm {eff}} = 0.16 \pm 0.02 $ with $m_{\tilde g} =$ 16 GeV.
Additional theoretical uncertainties
arise because there is no fully differential NLO calculation of gluino
production and subsequent decay to $b$'s.  The choice $m_{\tilde g} > 12$ GeV 
leads to a calculated $\bar{\chi}_{\rm {eff}}$ that is 
consistent with the data within experimental and theoretical uncertainties.  
With $\sigma_{\tilde{g}\tilde{g}} / \sigma_{\rm{qcd}} \sim 1/3$, 
the mixing data and the magnitude and $p_T$ dependence of the $b$ production 
cross section can be satisfied.  

\section{Constraints from other Data}
An early study by the UA1 Collaboration~\cite{ua1gluino} 
excludes $\tilde{g}$'s in the mass range $4 < m_{\tilde{g}} 
< 53$ GeV, but it starts from the assumption that there is a light neutralino 
${\tilde{\chi}}_1^0$ whose mass is less than the mass of the gluino.  The 
conclusion is based on the absence of the expected decay $\tilde{g} \rightarrow 
q+\bar{q}+\not{E_T}$, where $\not{E_T}$ represents the missing energy 
associated with the ${\tilde{\chi}}_1^0$.  In the scenario discussed above, this 
decay process does not occur since the bottom squark is the LSP, the SUSY particle 
with lowest mass, and the ${\tilde{\chi}}_1^0$ mass is presumed to be large 
({\em i.e.}, $> 50$ GeV). An analysis of 2- and 4-jet events by the ALEPH 
collaboration~\cite{ALEPH} disfavors $\tilde g$'s with mass 
$m_{\tilde g} < 6.3$ GeV but not $\tilde g$'s in the mass range relevant for the 
SUSY interpretation of the bottom quark production cross section.  A similar 
analysis is reported by the OPAL collaboration~\cite{OPALg}.  A light 
$\tilde{b}$ is not excluded by the ALEPH analysis.  
The exclusion by the CLEO collaboration~\cite{CLEO} of a $\tilde b$ with mass 
3.5 to 4.5 GeV does not apply since their analysis focuses only on the decays 
$\tilde b \rightarrow c \ell \tilde \nu$ and 
$\tilde b \rightarrow c \ell $.  The $\tilde b$ need not decay leptonically nor 
into charm.  On the other hand, these data might be 
reinterpreted in terms of a bound on the R-parity violating lepton-number 
violating decay of $\tilde b$ into $c \ell$.  It would be interesting  
to study the hadronic decays $\tilde b \rightarrow c q$, with $q = d$ or $s$, 
and $\tilde b \rightarrow u s$ with the CLEO data.  The DELPHI 
collaboration's~\cite{DELPHI} search for long-lived squarks in their 
$\gamma \gamma$ event sample is not sensitive to $m_{\tilde b} < 15$ GeV.  

There are important constraints on couplings of the bottom squarks from precise 
measurements of $Z^0$ decays.  A light $\tilde b$ would be ruled out unless its 
coupling to the $Z^0$ is very small.  The squark couplings to the $Z^0$ depend 
on the mixing angle $\theta_b$.  As described in Ref.~\cite{CHWW}, if the light 
bottom squark ($\widetilde{b}_1$) is an appropriate mixture of left-handed and 
right-handed bottom squarks, its lowest-order (tree-level) coupling to the $Z^0$ 
can be arranged to be small if $\sin^2 \theta_b \sim 1/6$.  The couplings 
$Z_{\tilde{b}_1 \tilde{b}_2}$ and $Z_{\tilde{b}_2 \tilde{b}_2}$ survive, where 
$\widetilde{b}_2$ is the heavier bottom squark.  However, as long as the 
combination of the masses $m_{\tilde{b}_1} + m_{\tilde{b}_2}$ is less than the 
maximum center-of-mass energy explored at LEP, these couplings present no 
difficulty.  This condition roughly implies $m_{\tilde{b}_2} > 200$ GeV.  However, 
much lower masses of $\tilde{b}_2$ might be tolerated.  A careful phenomenological 
analysis is needed of expected $\widetilde{b}_2$ decay signatures, along with 
an understanding of detection efficiencies and expected event rates, before one 
knows the admissible range of masses consistent with LEP data.  At higher-order, 
unless the $\widetilde{b}_2$ mass is of order 100 GeV, contributions from loop 
processes in which light gluinos are exchanged may produce significant deviations 
from measurements of the ratios $A_b$, the forward-backward $b$ asymmetry at the 
$Z^0$, and $R_b$, the hadronic branching ratio of the $Z^0$ into $b$ 
quarks~\cite{loops1,loops2}.    

Bottom squarks make a small contribution to the
inclusive cross section for $e^+ e^- \rightarrow$ hadrons, in comparison to
the contributions from quark production, and $\tilde{b} \bar{\tilde{b}}$
resonances are likely to be impossible to extract from backgrounds
\cite{Nappi}.  The angular distribution of hadronic jets produced in $e^+ e^-$ 
annihilation can be examined in order to bound the contribution of
scalar-quark production.  Spin-1/2 quarks and spin-0 squarks emerge with
different distributions, $(1 \pm {\rm cos}^2 \theta)$, respectively. 
The angular distribution measured by the CELLO
collaboration~\cite{CELLO} is consistent with the production
of a single pair of charge-1/3 squarks along with five flavors of
quark-antiquark pairs.  A new examination of the angular distribution 
with greater statistics would be valuable.

\section{Predictions and Implications}

\subsection{$\Upsilon$ Decay into Bottom Squarks}

If the bottom squark mass is less than half the mass of one of the Upsilon states,  
then Upsilon decay to a pair of bottom squarks might proceed with sufficient rate 
for experimental observation or exclusion of a light bottom squark.  In 
Ref.~\cite{elblc}, the expected rate for 
$\Upsilon \rightarrow \tilde b {\tilde b}^*$
is computed as a function of the masses of the bottom squark and the gluino.  The 
mass of the gluino enters because the $\tilde g$ is exchanged in the decay 
subprocesses.  The electronic width of the $\Upsilon$ is used as the source of 
absolute normalization.  

The data sample is largest at the $\Upsilon(4S)$.  For a fixed 
gluino mass of 14 GeV, the branching fraction into a pair of bottom 
squarks is about $10^{-3}$, for $m_{\tilde b} =$ 2.5 GeV, and about $10^{-4}$ for 
$m_{\tilde b} =$ 4.85 GeV.  A sample as large as $10,000$ may be available in 
current data from runs of the CLEO detector.  

The predicted decay rates for the $\Upsilon(nS)$, $n = 1$, 3 are shown in Fig. 2 
These curves can be read as predictions of the width for the corresponding values 
of $m_{\tilde b}$ and $m_{\tilde g}$, or as lower limits on the sparticle masses 
given known bounds on the branching fractions.  The current experimental 
uncertainties on the hadronic widths of the $\Upsilon$'s are compatible with the 
range of values of $m_{\tilde b}$ and $m_{\tilde g}$ favored in the work on the 
bottom quark production cross section in hadron reactions described above.  The 
analysis of $\Upsilon(nS)$ decays shows nevertheless that tighter experimental 
bounds on the bottom squark fraction are potentially powerful for the 
establishment of lower bounds on $m_{\tilde b}$ and $m_{\tilde g}$.   
\begin{figure}
\caption{Loci in the
$m_{\tilde{b}}$ -- $m_{\tilde{g}}$ plane for which the rate for Upsilon decay into 
a pair of bottom squarks is either 10 times $\Gamma_{\ell \bar{\ell}}$ for each  
$\Upsilon(nS)$, with $n= 1 - 3$ (lower set of curves) or equal to 
$\Gamma_{\ell \bar{\ell}}$ 
(upper set).  Above the curves, the rate would be less.  Within each set 
of curves, the order is (bottom to top) $1S, 2S, 3S$.}
\includegraphics[height=.48\textheight]{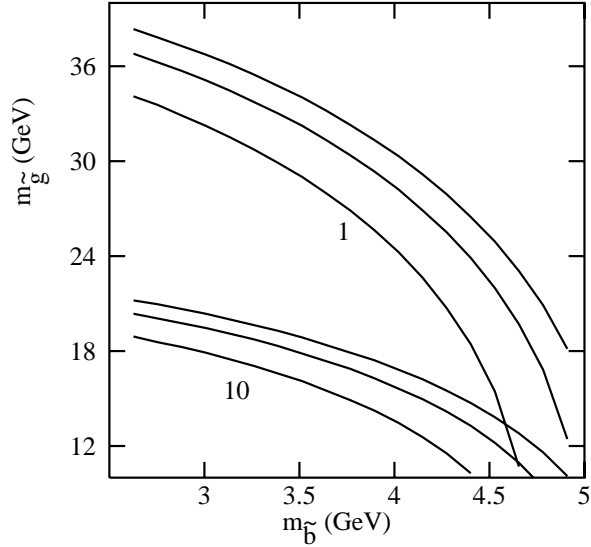}
\end{figure}

The hadronic width of the $\Upsilon$ is calculated in conventional QCD perturbation 
theory from the three-gluon decay subprocess, $\Upsilon \rightarrow 3g$, and 
$\Gamma_{3g} \propto \alpha^3_s$.  The SUSY 
subprocess adds a new term to the hadronic width from 
$\Upsilon \rightarrow \tilde{b} + \bar{\tilde{b}}$.  If this new subprocess is 
present but ignored in the analysis of the hadronic width, the true value of 
$\alpha_s(\mu = m_b)$ will be smaller than that extracted from a standard QCD 
fit by the factor $(1 - \Gamma_{\rm{SUSY}}/\Gamma_{3g})^{\frac{1}{3}}$.
For a contribution from the $\tilde{b} \bar{\tilde{b}}$ final state
that is 25\% of $\Gamma^{\Upsilon}(1S)$  ({\em i.e.}, a ratio 
$R^{\Upsilon}_{\tilde{b} \bar{\tilde{b}}} = 10$ in Fig.~2), the value of 
$\alpha_s$ extracted will be reduced by a factor of 0.9, at the 
lower edge of the approximately 10\% uncertainty band on the commonly quoted 
value of $\alpha_s(m_b)$~\cite{alfs}.   A thorough analysis 
would require the computation of next-to-leading order contributions in SUSY-QCD to 
both the $3g$ and $\tilde{b} \bar{\tilde{b}}$ amplitudes and the appropriate 
evolution of $\alpha_s(\mu)$ with inclusion of a light gluino and a light bottom 
squark.  

Direct observation of 
Upsilon decay into bottom squarks requires an understanding of the ways 
that bottom squarks may manifest themselves, discussed in more detail below.  
Possible baryon-number-violating R-parity-violating decays of the bottom squark 
lead to $u+s$; $c+d$; and $c+s$ final states.  These final states of four 
light quarks should be distinguishable from conventional hadronic final states 
mediated by the three-gluon intermediate state.  If the $\tilde{b}$ lives long 
enough, it will pick up a light quark and turn into a $B$-mesino, $\widetilde{B}$.
Charged $B$-mesino signatures in $\Upsilon$ decay include single back-to-back equal
momentum tracks in the center-of-mass; measurably lower momentum than 
lepton pairs ( $< 4$ GeV/c {\em{vs.}} $\simeq 5$ GeV/c for muons and electrons);
$1 + \cos^2 \theta$ angular distribution; and ionization, time-of-flight, 
and Cherenkov signatures 
consistent with a particle whose mass is heavier than that of a proton.  At stake 
is discovery, or new limits on the mass, of the ${\tilde b}$ as well as 
measurement of or new limits on the R-parity violation couplings of the 
${\tilde b}$.  

Pseudoscalar $\eta_b$ decay into a pair of bottom squarks is forbidden, but the 
higher-order process of $\eta_b$ decay into a pair of $B$-mesinos can proceed.
Decays of the $\chi_{b0}$ and $\chi_{b2}$ into a pair of bottom squarks are 
allowed.  

\subsection{Hadron Reactions}
Among the predictions of this SUSY scenario, the most clearcut is pair
production of like-sign charged $B$ mesons at hadron colliders, $B^+B^+$ 
and $B^-B^-$.  To verify the underlying premise, that the cross 
section exceeds expectations of conventional perturbative QCD, a new measurement 
of the absolute rate for $b$ production in run II of the Tevatron is important.   
A very precise measurement of $\bar{\chi}$ in run II is desirable.  
Since the fraction of $b$'s from gluinos changes with $p_{Tb}$, a change of 
$\bar{\chi}$ is expected when the cut on $p_{Tb}$ is changed.  The $b$ jet 
from $\tilde{g}$ decay into $b\widetilde{b}$ will 
contain the $\widetilde{b}$, implying unusual material associated with the 
$\widetilde{b}$ in some fraction of the $b\bar{b}$ data sample.  The existence 
of light $\widetilde b$'s means that they will be pair-produced in partonic 
processes, leading to a slight increase ($\sim 1$\%) 
in the hadronic dijet rate.  

The SUSY approach increases the $b$ production rate
at HERA and in $\gamma \gamma$ collisions at LEP by a small amount, not
enough perhaps if early experimental indications in these cases are
confirmed~\cite{herab,lepb}.  Full NLO SUSY-QCD studies should be undertaken. 
In these two cases, the apparent discrepancy may find at least part of its 
resolution in the fact that $b \bar{b}$ production occurs very near threshold 
where fixed-order QCD calculations are not obviously reliable.  Uncertain parton 
densities of photons may play a significant role.  

\subsection{Running of $\alpha_s$}
The presence of a light gluino and a light bottom squark slow the running of the 
strong coupling strength $\alpha_s(\mu)$. Above gluino threshold, the $\beta$ 
function of (SUSY) QCD is 
\begin{equation}
\beta(\alpha_S) = \frac{\alpha_S^2}{2 \pi} \left( -11 + \frac{2}{3} n_f + 
\frac{1}{6} n_s + 2 \right) .  
\end{equation}
The $\widetilde{b}$ (color triplet scalar) contributes little to the running, 
equivalent to that of 1/4'th of a new flavor,  
but the $\widetilde{g}$ (color octet fermion) is much more significant, equivalent 
to that of 3 new flavors of quarks.  A precise determination of 
$\beta(\alpha_S)$ appears to be the best way to exclude the presence of a light 
gluino or to establish its possible presence.   

In the standard model, a global fit to all observables provides an indirect 
measurement of $\alpha_s$ at the scale of the $Z$ boson mass $M_Z$.  The value 
$\alpha_s(M_Z) \simeq 0.1184 \pm 0.006$ describes most observables 
properly~\cite{alfs}.  Extrapolation from $M_Z$ to a lower scale $\mu$ with 
inclusion of a light gluino reduces $\alpha_s(\mu)$ from its pure QCD value.  
The presence of a light gluino, with or without a light bottom squark, also 
requires reanalysis of the phenomenological determinations of 
$\alpha_s(\mu)$ at all scales to take into account SUSY processes and SUSY-QCD 
corrections to the amplitudes that describe the relevant processes.  To date, a 
systematic study of this type has not been undertaken, but, as mentioned above, 
consistency is achieved for $\Upsilon$ decays.  A lesser value of $\alpha_s(m_b)$ 
leads, under slower evolution, to the same $\alpha_s(M_Z)$.  
  
Slower running of $\alpha_s(Q)$ also means a 
slower evolution of parton densities at small $x$, an effect that might be 
seen in HERA data for $Q > m_{\tilde{g}}$.  Presence of a scalar $\tilde{b}$ in 
the proton breaks the Callan-Gross relation and yields a non-zero leading-twist 
longitudinal structure function $F_L(x,Q)$ at leading-order.

\subsection{$\widetilde b$-onia}
Bound states of bottom squark pairs could be seen as $J^P = 0^+$, 
$1^-$, $2^+$, ... mesonic resonances in $\gamma \gamma$ reactions and in 
$p\bar{p}$ formation, with masses in the 4 to 10 GeV range.  They could 
show up as narrow states in the $\mu^+ \mu^-$ invariant mass spectra at hadron 
colliders, between the $J/\Psi$ and $\Upsilon$.  At an $e^+ e^-$ collider, the 
intermediate photon requires production of a $J^{PC} = 1^{--}$ state.  Bound 
states of low mass squarks with charge $2/3$ were studied with a potential 
model~\cite{Nappi}.  The small leptonic widths were found to preclude bounds 
for $m_{\widetilde{q}} > 3$ GeV.  For bottom squarks with charge $-1/3$, the 
situation is more difficult. 

\subsection{$\widetilde b$ lifetime and observability}
Strict R-parity conservation in the MSSM forbids $\widetilde{b}$ decay 
unless there is a lighter supersymmetric particle.  R-parity-violating and 
lepton-number-violating decay of the $\widetilde{b}$ into at least one lepton is 
disfavored by the CLEO data~\cite{CLEO} and would imply the presence of an extra 
lepton, albeit soft, in some fraction of $b$ jets observed at hadron colliders.  
The baryon-number-violating R-parity-violating (${\not \! R_p}$) term in the MSSM 
superpotential is 
${\cal W}_{\not \! R_p} = \lambda_{ijk}^{\prime\prime}U_i^cD_j^cD_k^c$; 
$U^c_i$ and $D^c_i$ are right-handed-quark singlet chiral
superfields; and $i,j,k$ are generation indices.  The limits on individual 
${\not \! R_p}$ and baryon-number violating couplings 
$\lambda''$ are relatively weak for third-generation squarks~\cite{Allanach,BHS}, 
$\lambda''_{ijk} < 0.5$ to $1$.  

The possible ${\not \! R_p}$ decay channels 
for the $\widetilde{b}$ are $123: \bar{\widetilde{b}} \rightarrow u+s$; 
$213: \bar{\widetilde{b}} \rightarrow c+d$; and 
$223: \bar{\widetilde{b}} \rightarrow c+s$.  The hadronic width is~\cite{BHS} 
\begin{equation}
\Gamma(\widetilde b \rightarrow \rm{jet} + \rm{jet}) =
\frac{m_{\widetilde b}}{2\pi} \sin^2\theta_{\widetilde{b}} 
\sum_{j<k} |\lambda^{\prime\prime}_{ij3}|^2 .
\end{equation}
If $m_{\tilde b} = 3.5$ GeV, $\Gamma(\widetilde b \rightarrow i j) = 
0.08 |\lambda^{\prime\prime}_{ij3}|^2$ GeV.  Unless all 
$\lambda^{\prime\prime}_{ij3}$ are extremely small, the $\widetilde {b}$ will 
decay quickly and leave soft jets in the cone around the $b$.  $b$-jets with an 
extra $c$ are possibly disfavored by CDF, but a detailed simulation is needed.

If the $\widetilde{b}$ is relatively stable, the $\widetilde{b}$ could pick 
up a light $\bar{u}$ or $\bar{d}$ and become a $\widetilde{B}^-$ or 
$\widetilde{B}^0$ ``mesino" with $J = 1/2$, the superpartner of the $B$ meson.  
The mass of the mesino would fall roughly in the range $3$ to $7$ GeV for 
the interval of $\widetilde{b}$ masses we consider.  The charged mesino could fake 
a heavy muon if its hadronic cross 
section is small and if it survives passage through the hadron calorimeter and exits 
the muon chambers.  Extra muon-like tracks would then appear in a fraction of the 
$b \bar{b}$ event sample, but tracks that leave some activity in the hadron calorimeter.  
The mesino has baryon number zero but acts like a 
heavy $\bar{p}$ -- perhaps detectable with a time-of-flight apparatus. 
A long-lived $\widetilde b$ is not excluded by conventional searches at hadron 
and lepton colliders, but an analysis~\cite{baeretal} similar to that for 
$\tilde{g}$'s  should be done to verify that there are no 
additional constraints on the allowed range of $\tilde b$ masses and lifetimes.
\begin{theacknowledgments}
I am indebted to Brian~Harris, David~E.~Kaplan, Zack~Sullivan, Tim~Tait, Carlos~
Wagner and Lou Clavelli for their collaboration and suggestions.  I have benefitted from 
valuable discussions with Barry Wicklund, Tom LeCompte, Bruce Berger, and Harry 
Lipkin.  The research reported here was supported by the U.S. Department of Energy 
under Contract W-31-109-ENG-38.  This paper was prepared at the invitation of the 
organizers of the 9th International Symposium on Heavy Flavor Physics, Caltech, 
Pasadena, CA, September 10 - 13, 2001.  I thank David Hitlin, Frank Porter, and 
the other members of the Local Organizing Committee for an excellent, timely, and 
most enjoyable meeting.  
\end{theacknowledgments}

% choose bibtex style depending on layout style and options used in
% sample:

\doingARLO[\bibliographystyle{aipproc}]
          {\ifthenelse{\equal{\AIPcitestyleselect}{num}}
             {\bibliographystyle{arlonum}}
             {\bibliographystyle{arlobib}}
          }

%\bibliography{elbwriteup}

\end{document}